# Towards merged-element transmons using silicon fins: the FinMET

A. Goswami,[1, *] A.P. McFadden,[2] T. Zhao,[2,4] H. Inbar,[3] J.T. Dong,[3] R. Zhao,[2, 4] C.R.H. McRae,[2, 4, 5] R.W. Simmonds,[2] C.J. Palmstrøm,[1, 3,] and D.P. Pappas[2]

[1]Electrical and Computer Engineering Department, University of California, Santa Barbara, Santa Barbara, CA 93106
[2]National Institute of Standards and Technology, Boulder, Colorado 80305, USA
[3]Materials Department, University of California, Santa Barbara, Santa Barbara, CA 93106
[4]Department of Physics, University of Colorado, Boulder, Colorado 80309, USA
[5]Department of Electrical, Computer, and Energy Engineering, University of Colorado, Boulder, Colorado 80309, USA

A merged-element transmon (MET) device, based on silicon (Si) fins, is proposed and the first steps to form such a "FinMET" are demonstrated. This new application of fin technology capitalizes on the anisotropic etch of Si(111) relative to Si(110) to define atomically flat, high aspect ratio Si tunnel barriers with epitaxial superconductor contacts on the parallel side-wall surfaces. This process circumvents the challenges associated with the growth of low-loss insulating barriers on lattice matched superconductors. By implementing low-loss, intrinsic float-zone Si as the barrier material rather than commonly used, potentially lossy AlO$_x$, the FinMET is expected to overcome problems with standard transmons by (1) reducing dielectric losses; (2) minimizing the formation of two-level system spectral features; (3) exhibiting greater control over barrier thickness and qubit frequency spread, especially when combined with commercial fin fabrication and atomic-layer digital etching; (4) potentially reducing the footprint by several orders of magnitude; and (5) allowing scalable fabrication. Here, as a first step to making such a device, the fabrication of Si fin capacitors on Si(110) substrates with shadow-deposited Al electrodes is demonstrated. These fin capacitors are then fabricated into lumped element resonator circuits and probed using low-temperature microwave measurements. Further thinning of silicon junctions towards the tunneling regime will enable the scalable fabrication of FinMET devices based on existing silicon technology, while simultaneously avoiding lossy amorphous dielectrics for the tunnel barriers.

## INTRODUCTION

The invention of the transmon qubit has fueled the rapid development of quantum-information research over the past decade [1], and landmark breakthroughs have been achieved with this technology[2]. Modern transmons are typically based on a small Al/AlO$_x$/Al tunnel junction formed by thermal oxidation of Al in parallel with a large shunt capacitor. A variety of methods exist for defining the Josephson junctions (JJs) that function as nonlinear inductances. These include Dolan bridges [3], the Manhattan shadow evaporation process[4] and overlap designs[5], [6].

However, the transmon qubits are difficult to scale for a couple of reasons. First, the associated shunt capacitors are typically defined as planar structures grown on a low-loss substrate. While very low-loss substrates (i.e., intrinsic, float-zone silicon (i-Si)) with loss tangent in the low $10^{-7}$ range can be obtained [7], it is well known that the interfaces and surfaces of the planar shunt capacitor participate significantly in the total loss [8], [9]. In general, while it has been observed that increasing the size of the shunt capacitor can dilute the high loss contributions [10], this results in very large structures, with dimensions of the order of 100s of micrometers. This is problematic for scaling up to systems with many qubits and warrants capacitors with lower form factors. Second, the frequency allocations of transmons using thermally oxidized aluminum have a significant spread [11]. While there have been advances on this front using post-processing, i.e. laser-annealing of individual devices [12], this remains a significant obstacle to large-scale integration of transmons. To this end, better control of the tunnel-barrier thickness and interfaces in the fabrication process is desired.

Recently, an alternative approach to scaling these circuits was demonstrated [13], i.e. the merged-element transmon (MET). The MET minimizes the transmon qubit size and radiation while providing an avenue to potentially reduce losses due to surfaces and interfaces. This design entails engineering the junction itself to satisfy the transmon requirements for frequency, anharmonicity, and charge noise by merging the external shunt capacitor and the JJ inductance into a single element. This design is constructed from a

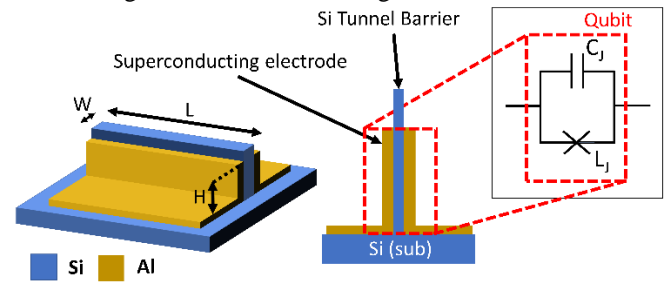

FIG 1 : 3D schematic of the FinMET and cross-section with corresponding circuit diagram of MET (inset).

superconductor–insulator-superconductor trilayer where the insulator is made from a dielectric material that has a low barrier height and may even be a semiconductor at room temperature. This design has several advantages over the traditional transmon. First, the MET allows a significant reduction, on the order of $10^4$, in the device area [13]. Second, the resulting small qubit dimensions effectively suppress unwanted radiation and qubit-qubit coupling through direct interactions or box modes. Third, the MET frequency should be less susceptible to the variation in lithography because the associated capacitive and inductive contributions toward the qubit frequency cancel out to first order [1], [13]. Moreover, one may choose a low-barrier-height material as the junction tunnel barrier. This enables the use of a relatively thick tunnel barrier that may reduce the percentage variation in junction inductance, potentially alleviating the frequency allocation problem.

The energy-level transitions of a MET device, from two-tone spectroscopy measurements confirmed that the MET is indeed operating in the transmon qubit regime [13]. An in-depth TLS-loss analysis identified the lossy amorphous silicon tunnel barrier and surrounding interfaces as the major limiting factor for the qubit relaxation time[13]. Subsequently, Mamin *et.al.* (and the IBM team) demonstrated METs with coherences up to 41 and 234 μs, using as-grown- and annealed-AlOx overlap junctions respectively [14].

While the MET demonstration [14] confirmed the possibility of obtaining long coherence times in selected devices, the extreme oxidation (several hours at 600 torr of $O_2$) and annealing conditions (rapid thermal anneal at 425°C) resulted in significant frequency spread for the devices. This is reminiscent of conventional transmons and is most likely due to several reasons. First, the tunneling critical current varies exponentially with the tunnel barrier thickness, which is difficult to control in a 2 nm thick tunnel barrier formed by thermal oxidization. In addition, there are tunneling hotspots due to the barrier being structurally and chemically inhomogeneous, resulting in only a small percentage of the 2 nm thick $AlO_x$ barrier actually contributing to the tunneling [15]. Second, the critical current can be affected by atomic-level defects in and around the barrier and wiring [16]. These can cause two-level-system spectral features[17] that are detrimental to the operation of the devices. This illustrates the importance of developing a more robust method of defining the tunnel junction. Specifically, we note that a crystalline tunnel barrier with a low barrier height can mitigate this problem because it can be thicker, making monolayer scale thickness variations less significant.

Here, we propose the concept of a FinMET device that can overcome a number of the problems discussed above. This process capitalizes on the fact that crystalline Si fins can be formed on the surface of a wafer using an anisotropic wet etch (FIG 1). In the FinMET. these Si fins act as tunnel barriers for the MET. The fin walls can, in principle, be atomically flat, parallel, and engineered to be a very specific thickness. With Si barriers the small band gap compared to that of $AlO_x$ also results in a lower tunnel-barrier height [13]

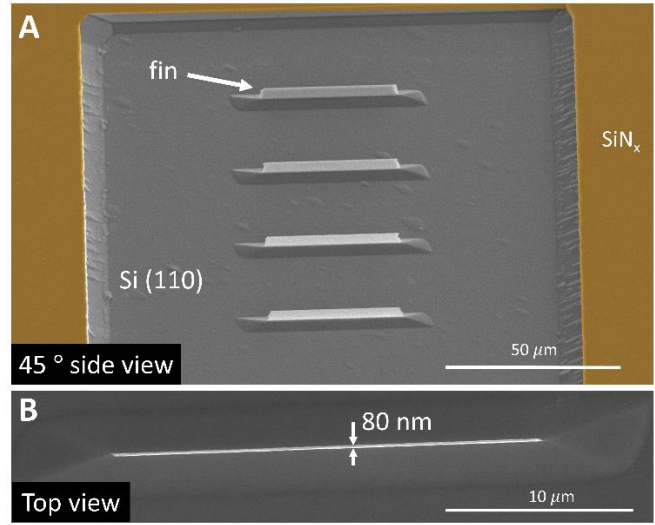

FIG 2 : Images of Si fins (A) 4 fins structures, $30\mu m$ long x 2.3 $\mu m$ tall and 80nm thick etched into a Si surface, including the trenched area around them (b) top view of a fin that is 30um long and 80nm thick.

. This allows the use of a relatively thick Si tunnel barrier, on the order of 5-10 nm, leading to a natural extension of the MET to a more scalable geometry.

In addition to being used as a junction, the fin geometry also allows the fabrication of low-loss parallel plate silicon capacitors with a significant reduction in utilized area on chip. Such capacitors can be used in conjunction with conventional Josephson junctions to reduce the size of conventional transmons while improving their coherence times or to enable new, novel protected qubits [18].

In this work, as first steps towards realizing a FinMET device, we demonstrate both the fabrication of high-aspect-ratio Si fins and the self-aligned process to deposit superconductors on such fins. Subsequently, we integrate these Al-Si-Al fin capacitors in resonator circuits and perform microwave measurements at millikelvin temperatures. The measurements clearly demonstrate the presence of working fin capacitors and their compatibility with traditional superconducting qubit fabrication. These scalable techniques are unique in that both capacitive elements and METs can be formed from the Si substrate resulting in reduction in both form-factor and losses and directly improving qubit scalability.

## FIN FABRICATION AND STRUCTURAL CHARACTERIZATION

Both fin-based capacitors and FinMET designs comprise parallel-plate electrodes that can be fabricated using standard processing techniques, and capacitively coupled through the low-loss substrate for scalable integrated circuits. In this case, it is critical to have low loss, epitaxial interfaces between the fin and the metal layer. This can be achieved using careful

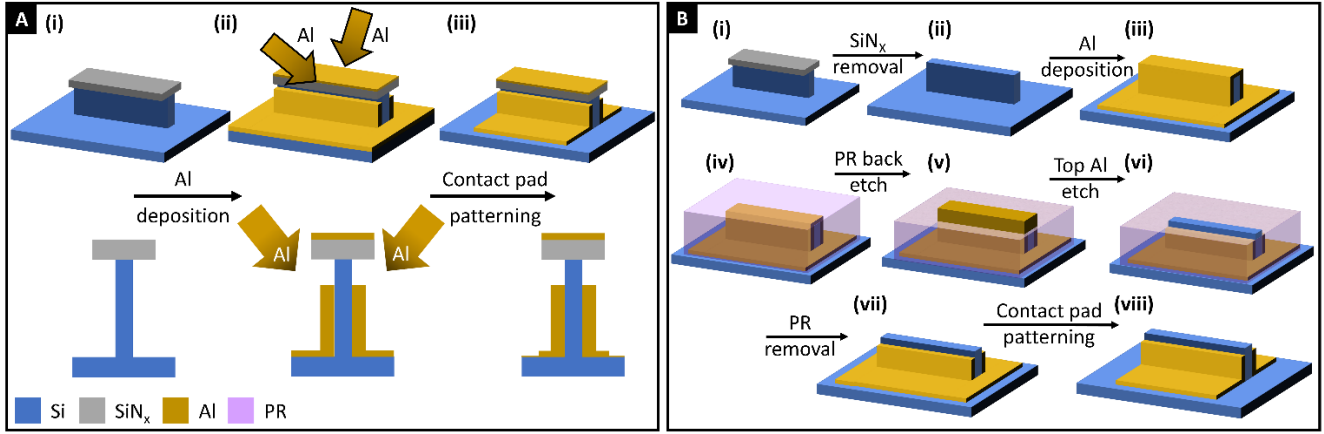

FIG 3 : Schematics of (A) self-aligned process and (B) proposed planarization process with process A being followed in this work.

surface cleaning and growth methods, as shown by Place, et al. [19].

**Fin fabrication** : The FinMET devices are composed of a Si fin [20], [21] with superconducting electrodes grown on both the surfaces of the fin, effectively forming a horizontal superconductor-semiconductor-superconductor junction, as illustrated in Figure 1. The Si fin is formed by top-down etching of a commercial intrinsic Si substrate, which are commonly grown using float-zone technique and exhibit high resistance and low loss. To achieve atomically flat surfaces, we start with a Si(110) substrate and use anisotropic wet etching to fabricate Si fins with smooth (111) surfaces.

For the anisotropic wet etch, a silicon nitride ($SiN_x$) hard mask is used. The $SiN_x$ layer is deposited using a low-pressure chemical vapor deposition technique which results in a high density, low stress $SiN_x$ layer on top of the silicon substrate. This mask is lithographically defined using electron beam lithography (EBL) and $CF_4/O_2$ based plasma dry etching.

Before the wet etch to define the fin, the sample is cleaned under a $C_4F_8/SF_6/CF_4$ Si etching chemistry in a plasma based dry etcher for 30 seconds to remove the top damaged layer from the previous dry etch for the $SiN_x$ mask. This was found to considerably improve the cleanliness of the substrate and the uniformity of the wet etch. The wet etch is subsequently performed using 45% potassium hydroxide (KOH) solution at 87°C. Figure 2 shows fin structures defined on a Si(110) substrate. Fin mask widths were varied from 100nm to 1um for test samples. The undercutting from the wet etch is considerable (~50-80nm). The thinnest fins that were reliably fabricated without rigorous optimization of process parameters were approximately 80nm (with a length and height of 100um and 3um respectively).

By using the intrinsically parallel crystallographic {111}-faces that are exposed by the anisotropic etch, the tunnel junction can be expected to have well-defined, homogeneous tunneling currents. In addition, the low-tunnel-barrier-height material can be much thicker than standard junction material and thus mass-fabricated with better margins.

**Fin metallization:** Metallization of the fins can be accomplished by cleaning the exposed Si{111}-faces and then epitaxially growing a superconducting metal, such as Al. Prior to Al growth, a buffered HF etch followed by high temperature annealing of the Si fin is expected to result in a pristine interface, thus optimizing coherence. Additionally, since the bottom edge is surrounded by low-loss substrate a reduced participation of the Al-air exposed interface is expected.

Two different process flows can be followed, either self-aligned or planarized, as illustrated in Figure 3. Both processes start by etching Si fins as described previously, using a $SiN_x$ hard mask and a combination of dry and wet etches. The planarization process is not demonstrated here but explained in the discussion section.

The process flow that is followed in this article, involves retention of the SiNx hard mask that was initially used to etch down the Si fins. Using the overhang in the SiNx on top of the fin, an angled deposition of aluminum can lead to a break in the metal layer at the top of the fin. This shadow-evaporation technique therefore enables direct creation of a fin capacitor using a self-aligned process. This eliminates the need to etch off Al from the top of the fin to electrically disconnect the pads on either side of the fin (Fig. 3A). Contact pads can then be patterned using optical lithography. This self-aligned process was followed in this work. Figure 4 shows fin structures coated with Al, as described below. Fins were fabricated according to the above recipe and the Al was shadow deposited with an effusion cell in a molecular beam epitaxy chamber. The substrate temperature was maintained close to 0°C (Fig. 4 A-F). Smoother Al films can be grown by performing the deposition in a cryogenic cold stage maintained at -196 °C following a process similar to Refs. [22]–[25]. For the microwave measurements, the Al layer on the fins was shadow deposited using an electron beam evaporator at room temperature.

Connection to other parts of the circuit, for example superconducting resonators and wiring, can be accomplished using the Al pads on the sides of the fins using standard processing techniques.

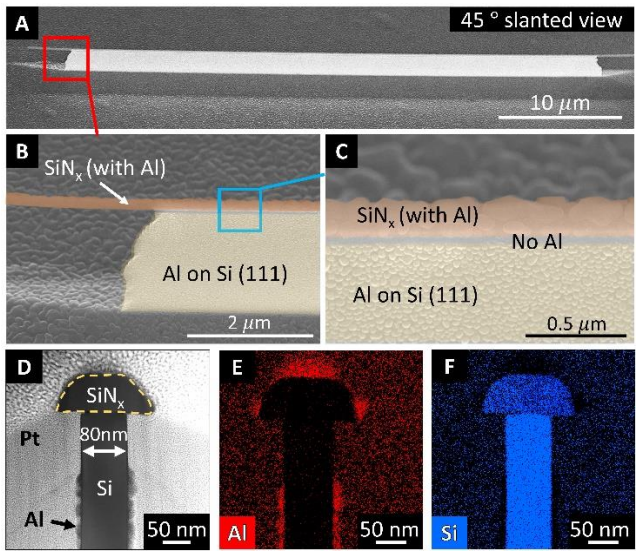

FIG 4: Metallized fin structure illustrating the self-aligned process and growth of Al on the Si{111} surfaces with a $SiN_x$ hard mask. (A) side view of the fin, (B) shows zoomed in area of (A) with the SiNx hard mask on top of the fin, extending out to the left, (C) shows a zoomed in area of (B) with the area where the Al is shadowed.(D) High angle dark field scanning transmission electron microscopy image of a fin cross-section with $SiN_x$ hard mask and shadow deposited Al. (E) and (F) shows energy-dispersive x-ray scans highlighting the Al and Si areas respectively

3um tall fins having lengths of 120um and patterned widths of 300nm and 400nm were fabricated on a 50.8mm diameter Si(110) wafer using the process outlined above. The fin capacitor height and width are set during fin formation while the fin capacitor length is lithographically defined in the metallization (superconductor) layer using optical lithography. Following fin formation, the wafer was etched in 6:1 BHF for 2 minutes to remove surface oxide and immediately loaded into high vacuum for Al deposition.

30nm of Al was deposited using e-beam evaporation on each side of the fins using deposition angles of $\pm 25^0$ to obtain a uniform Al coating on each side of the fins with a self-aligned break in the Al near the top of the fin due to the overhang of the silicon nitride mask. Following Al deposition, the test resonator circuits were defined using direct-write photolithography and wet etching. The wafer was diced into

## MICROWAVE MEASUREMENTS

Lumped element (LE) resonators were made using fin capacitors having widths greater than 200nm, meander inductors, and interdigitated capacitors (IDCs). These LE resonators were used to demonstrate feasibility of integration of fins in superconducting circuits and to characterize fabricated fins in the capacitive regime and evaluate preliminary dielectric losses. The concept of the LE design is depicted in Fig. 5. The design consists of eight LE resonators inductively coupled to a single 50 Ohm coplanar waveguide feedline in a standard 'hanger' arrangement[26].

To characterize the fin capacitors and minimize uncertainties due to capacitance of metallized leads, fins are incrementally added between the fingers of the IDC as shown in Fig. 5. The meander inductor and other IDC parameters are kept the same while the number of fins in each LE resonator is varied. In this way, the resonant frequencies of the LE resonators are separated, and capacitance of the fins may be determined experimentally.

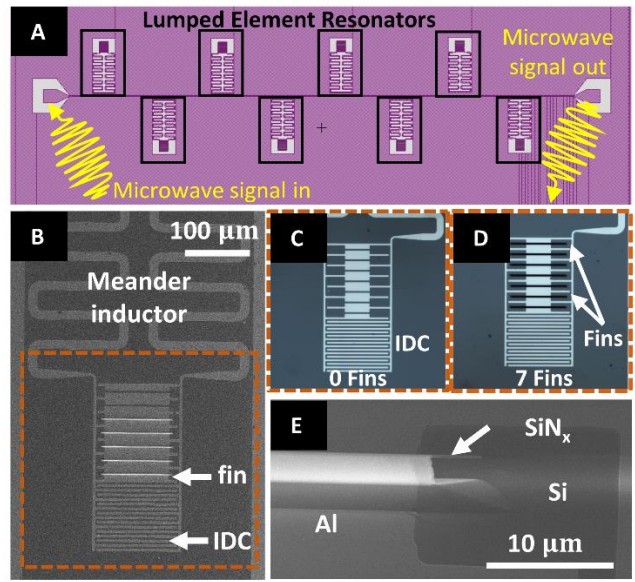

FIG 5 : (A) Mask layout for frequency multiplexed lumped element resonator (marked in black rectangles) design (B) SEM micrograph of a interdigitated capacitor (IDC) with fins coupled to a meander inductor forming the resonator (C) and (D) optical micrographs of two IDCs with 0 and 7 fins respectively (E) SEM micrograph showing aluminum patterned around 100 micron long fins.

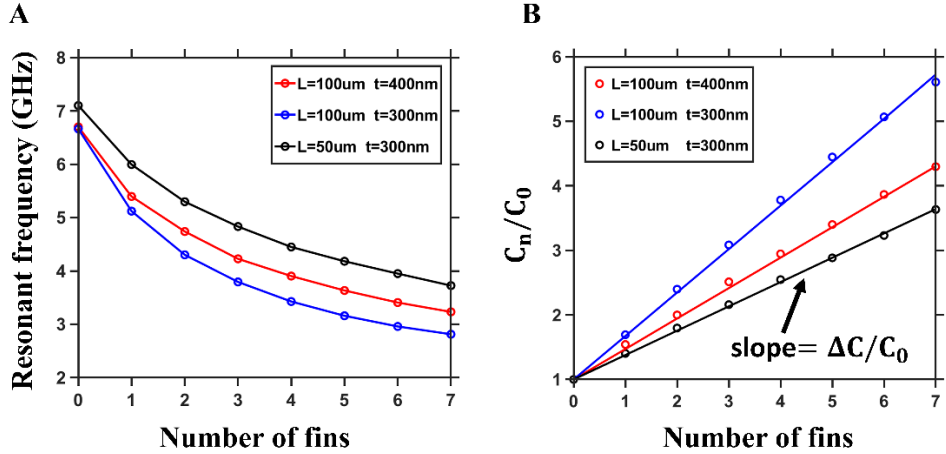

FIG 6: (A) Measured resonant frequency vs. number of fins in each interdigitated capacitor. (B) Capacitance ratio vs. number of fins (circles) along with fits (straight lines) from COMSOL simulations. Fits in (B) are for 219 nm and 319nm thicknesses for nominally 300nm and 400nm thick fins respectively.

7.5mm x 9.5mm die which were then packaged and cooled to 35mK for microwave measurements in a dilution refrigerator.

Figure 6(A) shows resonant frequency dependance on the number of fins included between the IDC fingers in each LE resonator. Three different die were measured where each die incorporates nominally identical fins and where the fin dimensions are varied between die. Figure 6(B) shows $C_n/C_0$ plotted vs number of fins incorporated in the IDCs for each of the three dies measured. Here $C_n$ is the capacitance of a LE resonator with n fins, and $C_0$ is the capacitance of the LE resonator with no fins. The linear trend in the data shown in Fig.6(B) is consistent with the expectation that each fin adds a set capacitance to the resonator, with the interpolated slope equal to $\Delta C/C_0$, where $\Delta C$ is the added capacitance from a single fin and $C_0$ is the capacitance of the LE resonator without any fins added.

We also simulated the capacitor part of the LE resonator with COMSOL® Electrostatics[27]. In these simulations, fin dimensions were chosen based on SEM micrographs. The simulation results were found to be in good agreement with experiment and are included in Fig. 6(B). The fin dimensions found to result in the best fit are indicated in Fig.6. For these fits, the fin capacitor length was kept constant at 100 μm (or 50 μm), the fin height was kept constant at 3.55 μm and the thickness was extracted from least square fitting (319nm for nominally 400nm thick fin and 219nm for nominally 300nm thick fin). For these dimensions, the fin capacitance per unit length along the fin was found to be ~1.3 fF/ μm for 319 nm wide fin and 1.8 fF/ μm for the 219nm wide fins. These values are in close agreement with the estimates obtained from a parallel plate geometry. In addition, the single-photon internal quality factors of one- and six-fins combined with the IDCs as shown in Fig.5 are determined to be $8.4 \times 10^4$ and $1.8 \times 10^5$ respectively. Following Ref. [28] and based on simulations of capacitance of the combined IDC-fin capacitors, we estimate the single-fin loss to be on the order of $\sim 3 \times 10^{-7}$. This is a promising demonstration of Si fins as an improvement over traditional thin-film capacitors and its potential development into FinMETs.

## DISCUSSION AND SUMMARY

The use of amorphous-$AlO_x$ Josephson junctions for quantum computing transmon applications is challenged by frequency allocation, frequency stability, spurious two-level states (TLS) and loss issues. In addition, the large size of shunt capacitors required to dilute the TLS losses at surfaces and interfaces severely limits the scalability.

In order to transition to a robust and more scalable technology, a significant effort on the front-end is required in order to bootstrap a completely new junction fabrication process and device design. To this end, we have proposed and demonstrated one of the necessary elements of a new FinMET technology. These include the fin structures needed to develop a more scalable system and reduce footprint. Although the fin thicknesses used in this work are not low enough to demonstrate tunneling, working low-loss fin capacitors integrated with superconductor resonator circuitry have been demonstrated and characterized. Next steps include integration of these capacitors with tunnel junctions to form qubits with reduced footprint and demonstration of tunnel junctions in the fins to form the proposed FinMET device.

To achieve tunneling through the silicon fins the fin thicknesses need to be approximately 5-10 nm. Structures of such extreme aspect ratios are on the cutting edge of modern fin technologies [20]. Thinning of the fins can then be achieved, by timing an additional wet etch and/or subsequent digital etching. The digital etching process is typically achieved by oxidizing the Si(111) surface using $O_2$ plasma at room temperature to form an oxide layer that is approximately 5-7nm thick. This oxide can then be etched away using HF and the process repeated to achieve the desired fin thickness. The second envisioned process to thin the fin involves using atomic layer etching (ALE) with a $O_2$, HF, and $Al(CH_3)_3$ chemistry [29]. A final wet etch in KOH can then be used to regain the smooth Si(111) surface followed by a HF dip to remove any oxides, prior to Al metal deposition.

The entire process can be also performed using photolithography by using $SiN_x$ deposited on step edges [30] that can form $SiN_x$ masks with similar dimensions. This can result in a more reliable wafer-scale fabrication of fin-based qubit devices.

To deposit the side superconductors, in addition to the process demonstrated, the second proposed process flow uses planarization (Fig. 3B). This involves removing the $SiN_x$ hard mask and then depositing a layer of aluminum onto the fins. Back-etching of a subsequent resist layer, either by dry etching or chemical mechanical polishing (CMP), can thereafter be used to expose the metal at the top of the fin and a wet or dry etch is used to remove the top aluminum. Contact pads would then be patterned in a way similar to the previous process flow.

While this is feasible to demonstrate at the small scale, the importance of these developments is that it should be possible to scale up significantly by decreasing the yield variation and material defects, based on existing Si-fin infrastructure and expertise in the field.

In conclusion, in this work, we first propose and outline a new geometry to define Si fin based self-aligned superconductor/Si/superconductor trilayer structures to form fin merged element transmons (FinMETs). Following this, we successfully fabricate, and measure low-loss resonators consisting of Al/Si/Al fin capacitors. This also establishes the compatibility of this technology with conventional qubit or novel [18]qubit fabrication platforms. Further thinning of the Si barrier layer will enable the subsequent realization of FinMET devices.

## DATA AVAILABILITY

The data that support the findings of this study are available from the corresponding author upon reasonable request.

## AUTHOR CONTRIBUTION

AG, APM, CJP and DPP initiated the study. AG fabricated the Si fin structures. AG, HI, JTD performed MBE Al deposition. APM performed electron beam Al deposition and fabrication of resonators. AG performed SEM and TEM measurements. APM performed microwave measurements. APM, CRHM and RWS analyzed the measurement data. RZ contributed to the initial MET simulations. TZ performed COMSOL simulations. AG wrote the manuscript with inputs from all the authors, under supervision of CJP, RWS and DPP.

## ACKNOWLEDGMENTS

We acknowledge the support of the NIST Quantum Initiative and the U.S. National Science Foundation (Grant No. 1839136) for the microwave measurements, the new and emerging qubit science and technology (NEQST) program initiated by the US Army Research Office (ARO) under Grant No. W911NF2210052 for the device fabrication, UCSB NSF Quantum Foundry through Q-AMASE-i program award number DMR-1906325 for the initial process developments, the epitaxial growth and microscopy studies. We also acknowledge the use of shared facilities of the UCSB MRSEC (NSF DMR 1720256) and the Nanotech UCSB Nanofabrication facility. We thank Dustin Hite and Mike Vissers for valuable feedback on the manuscript as well as Jim Beall for LPCVD nitride growth and technical assistance. This material is not subject to copyright protection within the United States.

## REFERENCES

[1] J. Koch *et al.*, "Charge-insensitive qubit design derived from the Cooper pair box," *Physical Review A - Atomic, Molecular, and Optical Physics*, vol. 76, no. 4, 2007, doi: 10.1103/PhysRevA.76.042319.

[2] F. Arute *et al.*, "Quantum supremacy using a programmable superconducting processor," *Nature*, vol. 574, no. 7779, pp. 505–510, Oct. 2019, doi: 10.1038/s41586-019-1666-5.

[3] G. J. Dolan, "Offset masks for lift-off photoprocessing," *Applied Physics Letters*, vol. 31, no. 5, 1977, doi: 10.1063/1.89690.

[4] M. v. Costache, G. Bridoux, I. Neumann, and S. O. Valenzuela, "Lateral metallic devices made by a multiangle shadow evaporation technique," *Journal of Vacuum Science & Technology B, Nanotechnology and Microelectronics: Materials, Processing, Measurement, and Phenomena*, vol. 30, no. 4, 2012, doi: 10.1116/1.4722982.

[5] X. Wu, J. L. Long, H. S. Ku, R. E. Lake, M. Bal, and D. P. Pappas, "Overlap junctions for high coherence superconducting qubits," *Applied Physics Letters*, vol. 111, no. 3, 2017, doi: 10.1063/1.4993937.

[6] A. Stehli *et al.*, "Coherent superconducting qubits from a subtractive junction fabrication process," *Applied Physics Letters*, vol. 117, no. 12, 2020, doi: 10.1063/5.0023533.

[7] W. Woods *et al.*, "Determining Interface Dielectric Losses in Superconducting Coplanar-Waveguide Resonators," *Physical Review Applied*, vol. 12, no. 1, 2019, doi: 10.1103/PhysRevApplied.12.014012.

[8] D. S. Wisbey *et al.*, "Effect of metal/substrate interfaces on radio-frequency loss in superconducting coplanar waveguides," *Journal of Applied Physics*, vol. 108, no. 9, 2010, doi: 10.1063/1.3499608.

[9] G. Calusine *et al.*, "Analysis and mitigation of interface losses in trenched superconducting

[10] J. M. Gambetta *et al.*, "Investigating surface loss effects in superconducting transmon qubits," *IEEE Transactions on Applied Superconductivity*, vol. 27, no. 1, 2017, doi: 10.1109/TASC.2016.2629670.

[11] J. M. Kreikebaum, K. P. O'Brien, A. Morvan, and I. Siddiqi, "Improving wafer-scale Josephson junction resistance variation in superconducting quantum coherent circuits," *Superconductor Science and Technology*, vol. 33, no. 6, 2020, doi: 10.1088/1361-6668/ab8617.

[12] J. B. Hertzberg *et al.*, "Laser-annealing Josephson junctions for yielding scaled-up superconducting quantum processors," *npj Quantum Information*, vol. 7, no. 1, 2021, doi: 10.1038/s41534-021-00464-5.

[13] R. Zhao *et al.*, "Merged-Element Transmon," *Physical Review Applied*, vol. 14, no. 6, 2020, doi: 10.1103/PhysRevApplied.14.064006.

[14] H. J. Mamin *et al.*, "Merged-Element Transmons: Design and Qubit Performance," *Physical Review Applied*, vol. 16, no. 2, 2021, doi: 10.1103/PhysRevApplied.16.024023.

[15] L. J. Zeng *et al.*, "Direct observation of the thickness distribution of ultra thin AlOx barriers in Al/AlOx/Al Josephson junctions," *Journal of Physics D: Applied Physics*, vol. 48, no. 39, 2015, doi: 10.1088/0022-3727/48/39/395308.

[16] S. Schlör *et al.*, "Correlating Decoherence in Transmon Qubits: Low Frequency Noise by Single Fluctuators," *Physical Review Letters*, vol. 123, no. 19, 2019, doi: 10.1103/PhysRevLett.123.190502.

[17] R. W. Simmonds, K. M. Lang, D. A. Hite, S. Nam, D. P. Pappas, and J. M. Martinis, "Decoherence in josephson phase qubits from junction resonators," *Physical Review Letters*, vol. 93, no. 7, p. 077003, Aug. 2004, doi: 10.1103/PHYSREVLETT.93.077003/FIGURES/4/MEDIUM.

[18] A. Gyenis *et al.*, "Experimental Realization of a Protected Superconducting Circuit Derived from the 0 - π Qubit," *PRX Quantum*, vol. 2, no. 1, p. 010339, Jan. 2021, doi: 10.1103/PRXQUANTUM.2.010339/FIGURES/16/MEDIUM.

[19] A. P. M. Place *et al.*, "New material platform for superconducting transmon qubits with coherence times exceeding 0.3 milliseconds," *Nature Communications*, vol. 12, no. 1, 2021, doi: 10.1038/s41467-021-22030-5.

[20] M. L. Chen *et al.*, "A FinFET with one atomic layer channel," *Nature Communications*, vol. 11, no. 1, 2020, doi: 10.1038/s41467-020-15096-0.

[21] P. S. Finnegan, A. E. Hollowell, C. L. Arrington, and A. L. Dagel, "High aspect ratio anisotropic silicon etching for x-ray phase contrast imaging grating fabrication," *Materials Science in Semiconductor Processing*, vol. 92, 2019, doi: 10.1016/j.mssp.2018.06.013.

[22] R. W. Simmonds *et al.*, "Josephson junction materials research using phase qubits," in *Quantum Computing in Solid State Systems*, 2006. doi: 10.1007/0-387-31143-2_11.

[23] B. M. McSkimming, A. Alexander, M. H. Samuels, B. Arey, I. Arslan, and C. J. K. Richardson, "Metamorphic growth of relaxed single crystalline aluminum on silicon (111)," *Journal of Vacuum Science & Technology A: Vacuum, Surfaces, and Films*, vol. 35, no. 2, 2017, doi: 10.1116/1.4971200.

[24] L. Aballe, C. Rogero, P. Kratzer, S. Gokhale, and K. Horn, "Probing interface electronic structure with overlayer quantum-well resonances: Al/Si(111)," *Physical Review Letters*, vol. 87, no. 15, 2001, doi: 10.1103/PhysRevLett.87.156801.

[25] H. Liu, Y. F. Zhang, D. Y. Wang, M. H. Pan, J. F. Jia, and Q. K. Xue, "Two-dimensional growth of Al films on Si(1 1 1)-7 × 7 at low-temperature," *Surface Science*, vol. 571, no. 1–3, 2004, doi: 10.1016/j.susc.2004.08.011.

[26] C. R. H. McRae *et al.*, "Materials loss measurements using superconducting microwave resonators," *Rev. Sci. Instrum. 91, 091101*, 2020, doi: 10.1063/5.0017378.

[27] "COMSOL Multiphysics® v. 6.0. www.comsol.com. COMSOL AB, Stockholm, Sweden".

[28] C. R. H. McRae *et al.*, "Dielectric loss extraction for superconducting microwave resonators," *Applied Physics Letters*, vol. 116, no. 19, May 2020, doi: 10.1063/5.0004622.

[29] A. I. Abdulagatov and S. M. George, "Thermal Atomic Layer Etching of Silicon Using O2, HF, and Al(CH3)3 as the Reactants," *Chemistry of Materials*, vol. 30, no. 23, 2018, doi: 10.1021/acs.chemmater.8b02745.

[30] V. Jovanovic, T. Suligoj, and L. Nanver, "Crystallographic Silicon-Etching for Ultra-High Aspect-Ratio FinFET," *ECS Transactions*, vol. 13, no. 1, 2008, doi: 10.1149/1.2911512.